\documentstyle[sprocl]{article}

\def\be{\begin{equation}}
\def\ee{\end{equation}}
\def\bea{\begin{eqnarray}}
\def\eea{\end{eqnarray}}

\begin{document}


\title{POLYAKOV CONJECTURE AND 2+1 DIMENSIONAL
GRAVITY\footnote{Presented by P. Menotti}} 

\author{L. CANTINI}

\address{Scuola Normale Superiore, Pisa and INFN Sezione di Pisa, Italy} 

\author{P. MENOTTI}

\address{Dipartimento di Fisica della Universit\`a, Pisa  and INFN
Sezione di Pisa, Italy}

\author{D. SEMINARA}

\address{Dipartimento di Fisica della Universit\`a, Firenze and INFN
Sezione di Firenze, Italy} 


\maketitle\abstracts{ After briefly reviewing the hamiltonian approach
to 2+1 dimensional gravity in absence of matter on closed universes,
we consider 2+1 dimensional gravity coupled to point particles in an
open universe. We show that the hamiltonian structure of the theory is
the result of a conjecture put forward by Polyakov in a different
context. A proof is given of such a conjecture. Finally we give the
exact quantization of the two particle problem in open space.}

\section{Introduction}

Gravity in $2+1$ dimensions \cite{DJH} has attracted considerable
interest in the 
past few years. Solutions in $2+1$ dimensions are also solution of
$3+1$ dimensional gravity in presence of a space-like Killing vector
field and for this reason they describe cosmic strings in some
special configuration.

The simplifying feature of gravity in $2+1$ dimensions is that the
Riemann tensor is a linear function of the Ricci tensor 
\begin{equation}
R^{\mu\nu}_{~~\lambda\rho} =
\epsilon^{\mu\nu\kappa}\epsilon_{\lambda\rho\sigma}
(R^\sigma_\kappa-\frac{\delta^\sigma_\kappa}{2} R) 
\end{equation}
and as such it vanishes outside the sources; there are no
gravitational waves 
in $2+1$  dimensions. Despite that, the theory is far from trivial
especially if we couple it to matter, and it retains several features
of the theory in higher dimensions. 

We shall see in this paper how the theory is strictly related to very
interesting mathematical structures, like Liouville theory and the
Riemann- Hilbert problem. The hamiltonian structure of
the theory will be related to a conjecture put forward by Polyakov
in the context of two dimensional quantum gravity and 
we shall give a proof of such a conjecture. It is possible to solve
explicitly the
quantum problem in some special cases and this results can throw light
in the far more complicated $3+1$ dimensional case.
 
In addition to the hamiltonian approach there are several other
approaches to the problem \cite{hooft,wael,nelsonregge,matschull}.
In this paper we shall summarize some old results and some recent
developments with regard to the hamiltonian formulation of the
theory. We shall insist on the conceptual side so that the paper can
work as a simple introduction to the subject; the technical details
can be found in the published papers and reports.

\section{Gravity in $2+1$ dimensions in absence of matter}
In absence of matter the only degrees of freedom are given by the
moduli of the space sections; thus we can deal only with genus $g\geq 1$.
 
In absence of boundaries the action of the gravitational field reduces
to the Einstein- Hilbert term which can be put in hamiltonian form as
\begin{equation}\label{hilbertaction}
S_H = \int dt \int_{\Sigma_t} d^Dx
\left[ \pi^{ij}\dot g_{ij} - N^i H_i - NH\right]
\end{equation}
where we used the standard ADM metric \cite{adm}
\begin{equation}
ds^2= -N^2 dt^2+ 2 g_{z\bar z}(dz+ N^z dt)(d\bar z+ N^{\bar z} dt).
\end{equation}

The choice of the gauge is of crucial importance in dealing with the
problem. The well known York gauge in which the time slices are
provided by the $D$ (in our case $D=2$) dimensional surfaces with $K=
{\rm const.}$, being $K$ the trace of the extrinsic curvature tensor
is particularly powerful as this gauge decouples the solution of the
diffeomorphism constraint from the hamiltonian constraint. Exploiting
this feature it is possible \cite{moncrief,hosoya} to solve the
diffeomorphism constraint and to provide the hamiltonian on the
reduced phase space given by the moduli and their conjugate
momenta. The number of moduli are $6g-6$ for genus $g$ larger than $1$
and $2$ for the torus topology. The explicit computation of the
hamiltonian can be performed only in the simplest case of torus
topology. 
It is given by 
\begin{equation}\label{torushamiltonian}
H = \sqrt{\tau^2_2(p_1^2+p_2^2)},  
\end{equation}
where the $p_j$ are the momenta conjugate to the moduli $\tau_j$.
Hamilton equations for the hamiltonian eq.(\ref{torushamiltonian}) can
be exactly 
solved \cite{hosoya,carlip} to obtain
$$
\tau_1 = c + a \tanh(t-t_0)
$$
\begin{equation}
\tau_2 = \frac{a}{\cosh(t-t_0)}
\end{equation} 
i.e. in the coordinate space the point moves along a semicircle lying
in the upper half plane with diameter on the real axis; such
semicircles are the geodesic of the Poincar\'e half- plane.
Quantization proceeds
\cite{carlip,hosoya2} by replacing the canonical variables with
operators according to the correspondence principle. The ordering
problem always subsists; the most natural ordering translates the
classical hamiltonian into the square root of the Maass laplacian
\cite{maass} giving rise to the Schroedinger equation 
\begin{equation}
i\frac{\partial\psi(x,y,t) }{\partial t}= \sqrt{-y^2(\partial_x^2+
\partial_y^2)}~\psi(x,y,t).  
\end{equation} 
The Maass laplacian has
been widely investigated by mathematicians
\cite{fay,terras,puzio}. The classical as well the quantum
hamiltonians are invariant under modular transformations which in the
case of the torus are given by the group $SL(2,Z)$ and thus the
solutions should be invariant under such modular transformations. The
eigenvalue problem under this condition is not trivial; nevertheless
it has been thoroughly studied, the general properties of the spectrum
is known even if it is not known in closed analytical form and the
eigenfunctions  cannot be simply expressed in terms of familiar
functions. Such an approach gives
a complete quantum treatment of universes without matter with torus
topology in the York gauge. Higher genuses are difficult to treat as
we do not know an explicit form of the hamiltonian for $g\geq 2$ and
probably here one will be able to assert only qualitative features of the
problem.
 
\section{Gravity in $2+1$ dimensions in presence of matter}

Obviously one can consider the case of closed and open universes. The
case of closed universes in presence of matter appears more difficult
to attack as in addition to the particle positions and
their conjugate momenta we have to deal with the moduli of the space
surface and their conjugate momenta. This however does not appear to be
the main difficulty. The favorable feature of working in an open
universe is that here the maximally slicing gauge (or Dirac gauge) can
be employed, which simplifies notably both the diffeomorphism and the
hamiltonian constraint. 

In presence of point particles the action has to be supplemented by
the matter action 
\begin{equation} 
S_m=\int\!d t \sum_n\Big(P_{ni}\, \dot
q_n^i+N^i(q_n) P_{ni} - N(q_n) \sqrt{P_{ni} P_{nj} g^{ij}(q_n)+
m_n^2}\Big)
\end{equation}
and for open universes by the boundary terms which play a fundamental
role. 

These are given by \cite{hawkinghunter}
\begin{equation}\label{boundaryaction} 
S_B = - \int dt H_B 
\end{equation} 
with
$$
-H_B= 2 \int_{B_t} d^{(D-1)}x \,\sqrt{\sigma_{Bt}} N 
\left( K_{B_t}+\frac{\eta}{\cosh\eta}{\cal D}_\alpha v^\alpha\right)
$$
\begin{equation}\label{boundaryhamiltonian} 
~~~~~~~~~~~~~~~-2 \int_{B_t} d^{(D-1)}x\, r_\alpha
\pi^{\alpha\beta}_{(\sigma_{Bt})} N_\beta. 
\end{equation}
The really important term turns out to be the first i.e. 
$\sqrt{\sigma_{Bt}} N K_{B_t}$ where  $K_{B_t}$ is the
extrinsic curvature of the $D-1$ dimensional boundary  (in our case
one- dimensional) of the time
slices as a sub-manifold embedded in the
$D$ dimensional time slices and $d^{(D-1)}x \sqrt{\sigma_{Bt}}$ is the
volume form induced by the space metric on the $D-1$ dimensional
boundary.  

We recall that the action $S_H+ S_B + S_m$ is so constructed as to provide
the correct 
equations of motion (i.e. Einstein's equations) when one computes a
stationary point of the action by keeping the values of the metric
fixed on the boundary. Such a procedure is equivalent \cite{wald} to the
weaker requirement of keeping fixed the intrinsic metric of the boundary.

One could in principle adopt also here the York
slicing. However the equations for the diffeomorphism and hamiltonian
constraints are not at all trivial. In particular the hamiltonian
constraint gives rise to an equation more complex than the
inhomogeneous sinh-Gordon equation. Progress has been
achieved by the introduction of the instantaneous York gauge
\cite{BCV,welling,MS1,MS2,CMS}, or maximally slicing gauge. This is defined by
all time slices having  
$K=0$. Simple application of the Gauss- Bonnet theorem shows that such
a gauge can be applied only to open universes, or universes with the
topology of the sphere \cite{welling,MS2}.  A closer
inspection shows that  
for the sphere topology such a gauge can describe only the simple
stationary 
case \cite{MS2}. Thus application of the $K=0$ gauge is practically
restricted 
to open universes, but here it proves very powerful. The technical
reason is the immediate solution of the diffeomorphism constraint 
given by
\begin{equation}\label{pibarzz}
\pi^{\bar z}_{~z} = -\frac{1}{2\pi}\sum_n\frac{P_n}{z-z_n}
\equiv -2\frac{\prod_B(z-z_B)}{{\cal P}(z)},
\end{equation}
where $z_n$ and $P_n$ are the complex positions and canonical
momenta of the particles. The $P_n$ are subject to the constraint
$\sum_n P_n=0$ \cite{Des85} which is related to the fact that
translations are not  symmetries of the problem but simply gauge
transformations \cite{Henn84,ashtekar}. The hamiltonian constraint
reduces to an inhomogeneous 
Liouville equation, given by
\begin{equation}\label{liouvilleeq}
2\Delta\tilde\sigma=-e^{-2\tilde\sigma}-4\pi \sum_n \delta^2(z- z_n)(
\mu _n -1)-4\pi\sum_B \delta^2(z- z_B),    
\end{equation}
to which powerful mathematical methods apply. 
Here $\tilde\sigma$ is defined by
\begin{equation}\label{sigma}
e^{2\sigma} = 2 \pi^{\bar z}_{~z} \pi^{z}_{~\bar z} e^{2\tilde\sigma},
\end{equation} 
and is related to the space metric by $g_{ij}= \delta_{ij}e^{2\sigma}$.
In the above equation the sources are given by the particles and in
addition by the zeros of eq.(\ref{pibarzz}) here denoted by
$z_B$; due to the constraint $\sum_n P_n=0$ they are $N-2$ in
number. They are know in the mathematical literature as apparent 
singularities. The strength of the particle sources are given by the
particle rest masses and that of the apparent singularities by the
constant $4\pi$. 
We shall rewrite eq.(\ref{liouvilleeq}) more generally as follows
\begin{equation}\label{pic}
4 \partial_z \partial_{\bar z}\phi = e^{\phi} +4\pi\sum_n
g_n\delta^2(z-z_n).   
\end{equation}
In a series of papers at the turn of the past century Picard
\cite{picard} proved that eq.(\ref{pic}) for real $\phi$
with asymptotic behavior at infinity
\begin{equation}
\phi(z) = -g_\infty\ln(z\bar z) + O(1)
\end{equation}
and $-1<g_n,~~1<g_\infty$ (which excludes the case of punctures) and
$\sum_n g_n +g_\infty < 0$
admits one and only one solution (see also \cite{poincare}). Picard
\cite{picard} achieved the solution of (\ref{pic}) through an
iteration process exploiting Schwarz alternating procedure. The same
problem has been considered recently with modern variational
techniques by Troyanov \cite{troyanov}, obtaining results which
include Picard's findings.

The above inequalities on the $g_n$ are all satisfied in gravity with
the following meaning: $1+g_n= \mu_n = m_n/4\pi$ being $m_n$ the rest
mass of the $n$-th particle which is constrained in the limits
$0<m_n<4\pi$. For $m_n>4\pi$ the conical defect exceeds its  maximum
value. $2-g_\infty = \mu = M/4\pi$ being $M$ the total energy of the
system which is also constrained to $0<M<4\pi$. 

From eq.(\ref{pic}) one can easily prove
\cite{poincare,bilalgervais} that the function $Q(z)$ defined by
\begin{equation}\label{bg}
e^{\frac{\phi}{2}} \partial_z^2 e^{-\frac{\phi}{2}} = -Q(z)
\end{equation}
is analytic i.e. 
$Q(z)$ is given by the analytic component of the energy momentum
tensor of a Liouville 
theory. $Q(z)$ is meromorphic with poles
up to the second order \cite{hempel} i.e. of the form 
\begin{equation}\label{Q}
Q(z) = \sum_n - \frac{g_n(g_n+2)}{4(z-z_n)^2} +
\frac{\beta_n}{2(z-z_n)}. 
\end{equation}
All  solutions of eq.(\ref{pic}) can be put in the form
\begin{equation}\label{mapping}
e^{\phi}=\frac{8f'\bar{f'}}{(1-f\bar f)^2} = \frac{8
|w_{12}|^2}{(y_2\bar y_2 - y_1\bar y_1)^2},~~~~
f(z) = \frac{y_1}{y_2}
\end{equation}
being $y_1,y_2$ two properly chosen, linearly independent solutions
of the fuchsian equation 
\begin{equation}\label{fuchs}
y''+Q(z)y=0.
\end{equation}
$w_{12}$ is the constant wronskian. 
This is a variant of the Riemann-Hilbert problem as we are not given
directly with the monodromies but with the following information on
them: all monodromies belong to the $SU(1,1)$ group, otherwise the conformal
factor (\ref{mapping}) is not single valued and as such cannot  
solve the Liouville equation eq.(\ref{pic}); in addition we
are given with the conjugation classes of the monodromies around 
the singularities and the conjugation class of the
monodromy at infinity.

The conformal factor $\tilde\sigma = -\phi/2$ is the
key quantity in all the 
subsequent developments. In fact a secondary constraint following from
the primary gauge constraint $K=0$ is
\begin{equation}\label{Nequation}
\Delta N = e^{-2\tilde\sigma} N
\end{equation}
and such $N$ can be computed from the knowledge of $\tilde
\sigma$. The reason is that the solution of the inhomogeneous Liouville
equation (\ref{liouvilleeq}) contains the free parameter $\mu$ which is
related to the  
behavior at infinity of the conformal factor i.e. $\exp(2\sigma) \approx s^2
(z\bar z)^{-\mu}$. As the sources do not depend on $\mu=M/4\pi$
a solution of
eq.(\ref{Nequation}) is given by
\begin{equation}\label{N}
N = \frac{\partial(-2\tilde\sigma)}{\partial M}
\end{equation}
and one easily proves such a solution to be unique \cite{MS1}.
The other secondary constraint on $N^{z}$ is 
\begin{equation}
\partial_{\bar z}N^z=- \pi^z_{~\bar z}~e^{-2\sigma} N
\end{equation}
solved by
\begin{equation}\label{Nz}
N^z =-\frac{2}{\pi^{\bar z}_{\ z}(z)} \partial_z N +g(z).
\end{equation}
Here $g(z)$ is a meromorphic function whose role is to cancel the
poles occurring in
the first member on the r.h.s. due to the
zeros of $\pi^{\bar z}_{\ z}$.
The expression of $g(z)$ in $N^z$ is \cite{MS1}
\begin{equation}\label{generalg}
g(z) = \sum_B \frac{\partial\beta_B}{\partial M} \frac{1}{z-z_B} \frac{{\cal
P}(z_B)}{\prod _{C\neq B} (z_B-z_C)} + p_1(z). 
\end{equation}
$p_1(z)$ is a first order polynomial related to the motion of
the frame at infinity, more properly to the
transformation  
of translations, rotations and dilatations which leave invariant the
conformal structure of the space metric and leave fixed the point
at infinity. In the following we shall examine the case $p_1(z)=0$;
this is no limitation as the two structures are related by a
canonical transformation \cite{CMS}. 
In conclusion the metric is obtained in a straightforward way from
$\tilde\sigma$. 

The equations of motion are extracted  from the action; for 
$p_1(z)=0$ they can be written as \cite{CMS}
\begin{equation}\label{equationsofmotion}
\dot z'_n = -\sum_B \frac{\partial \beta_B}{\partial
\mu}\frac{\partial z'_B}{\partial P'_n}~~~~{\rm and}~~~~
\dot P'_n = \frac{\partial \beta_n}{\partial \mu} +\sum_B
\frac{\partial \beta_B}{\partial
\mu}\frac{\partial z'_B}{\partial z'_n}
\end{equation}
where $z'_n = z_n-z_1$ and $P'_n =P_n$ for $n=2\dots N$.

The problem is now to show that such system is of hamiltonian nature
and possibly to write down the hamiltonian. The hamiltonian nature of
eqs.(\ref{equationsofmotion}) is expected as we obtained the above
equations by 
reduction of a hamiltonian system. Despite that, it is of interest to
have a direct proof of it and an expression of the hamiltonian.

In the simpler case of three body, where we have a single
accessory singularity one can prove the hamiltonian nature of the
equations of motion by exploiting Garnier equations which give a 
constraint on the evolution of the
accessory parameters under isomonodromic 
deformations \cite{CMS,yoshida}. The fact that our deformations
are isomonodromic is a 
consequence of the constancy in $2+1$ dimensional gravity of
the monodromies around the particle world lines. 

The problem with four or more particles, when we are in presence of two
or more accessory singularities is more difficult and it is related to an
interesting conjecture due to Polyakov \cite{conj} about the accessory
parameters 
of the $SU(1,1)$ Riemann-Hilbert problem. Such a conjecture states
that  the regularized Liouville action \cite{takh1,takh2}   
$$
S_\epsilon [\phi] =\frac{i}{2} \int_{X_\epsilon} (\partial_z\phi 
\partial_{\bar z} \phi +\frac{e^\phi}{2}) dz\wedge d\bar z
+\frac{i}{2}\sum_n g_n\oint_n\phi(\frac{d\bar z}{\bar z -\bar
z_n}- \frac{d z}{ z - z_n})
$$
\begin{equation}\label{Sepsilon}
+\frac{i}{2}g_\infty\oint_\infty\phi(\frac{d\bar z}{\bar z}- \frac{d
z}{z})
-\pi\sum_n g_n^2 \ln\epsilon^2 -\pi g_\infty^2\ln\epsilon^2 
\end{equation}
computed on the solution of the Liouville equation eq.(\ref{pic}) is the
generating function of the accessory parameters in the sense that
\begin{equation}
\beta_n = -\frac{1}{2\pi}\frac{\partial S_P}{\partial z_n}
\end{equation}
where $S_P = \lim_{\epsilon\rightarrow 0} S_\epsilon$.
Polyakov conjecture originated in the context of 2- dimensional
quantum gravity. Let $T(z)$ be the analytic component $T_{zz}$ of the
energy momentum tensor and let $V_\alpha$ be a primary field of weights
($\Delta_\alpha ,\Delta_{\bar\alpha}$) i.e. such that under a dilatation
$V_\alpha$ transforms like
\begin{equation}
V'_\alpha(kz) = k^{-\Delta_\alpha} k^{-\Delta_{\bar\alpha}} V_\alpha(z). 
\end{equation}
The Ward identity reads as
\begin{equation}
T(z)V_\alpha(w)
=\frac{\Delta_\alpha}{(z-w)^2}V_{\alpha}(w) +
\frac{1}{(z-w)}\partial_w V_{\alpha}(w) +\cdots  
\end{equation}
Given the field $V_{\alpha} = e^{\alpha \phi}$ with classical
conformal weights $(\alpha,\alpha)$ we have \cite{takh1,takh2} 
\begin{equation}
\langle V_{\alpha_1}(z_1)V_{\alpha_2}(z_2)\dots
V_{\alpha_N}(z_N)\rangle = \int D[\phi] e^{-\frac{S_P}{2\pi h}}  
\end{equation}
where $S_P$ is given by eq.(\ref{Sepsilon}) in the limit
$\epsilon\rightarrow 0$,  with $g_n= -h\alpha_n$. 
From the three point function one finds for the quantum weights
\begin{equation}
\Delta_{\alpha_n} = -\frac{h\alpha^2_n}{2} +\alpha_n =
\frac{1-\mu_n^2}{2h} 
\end{equation}
and
\begin{equation}
\langle T(z) V_{\alpha_1}(z_1)V_{\alpha_2}(z_2)\dots
V_{\alpha_N}(z_N)\rangle = \int D[\phi] \frac{1}{h}(\partial^2\phi 
-\frac{1}{2}(\partial\phi)^2) e^{-\frac{S_P}{2\pi h}}.
\end{equation}
In the classical limit i.e. for $h\rightarrow 0$, $\phi$ becomes the
solution of Liouville equation (\ref{pic}), related to the fuchsian
equation (\ref{fuchs}). 
But
\begin{equation}
e^{\frac{\phi}{2}}\partial_z^2 e^{-\frac{\phi}{2}} = -
\frac{1}{2}\partial^2 \phi+\frac{1}{4} 
(\partial\phi)^2 = -Q(z)
\end{equation}
which inserted in the previous equation gives again
\begin{equation}
\Delta_{\alpha_n} = \frac{1-\mu_n^2}{2h}
\end{equation}
and the new result
\begin{equation}\label{conjecture}
\beta_n = -\frac{1}{2\pi}\frac{\partial S_P}{\partial z_n}
\end{equation}
which is the content of Polyakov conjecture.

Zograf and Takhtajan \cite{ZT1} provided a proof of
eq.(\ref{conjecture}) for parabolic singularities using the technique
of mapping the quotient of the upper half-plane by a fuchsian group to
the Riemann surface and exploiting certain properties of the harmonic
Beltrami differentials. In addition they remark that the same
technique can be applied when some of the singularities are elliptic
of finite order. The case of only parabolic singularities is of
importance in the quantum Liouville theory \cite{takh2} as such
singularities 
provide the sources from which to compute the correlation
functions. On the other hand in $2+1$ gravity one is faced with
general elliptic singularities and here the mapping technique cannot
be applied. 
As a matter of fact we shall see that the case of elliptic
singularities is more closely related to the theory of elliptic non
linear differential equations (potential theory) than to the theory of
fuchsian groups.

We shall now briefly outline a derivation of Polyakov conjecture for
general elliptic singularities. 

We define $s_\infty^2$ and $s_n^2$ as the constant coefficients in the
asymptotic 
expansion of $\phi$ at the singularities i.e. 
\begin{equation}
\phi= - g_\infty \ln(z\bar z) -\ln s_\infty^2 + O(\frac{1}{|z|})
\end{equation} 
and 
\begin{equation}
\phi= g_n \ln((z-z_n)(\bar z-\bar z_n)) -\ln s_n^2 + O(|z-z_n|).
\end{equation}  
One then considers the derivative with respect to $z_n$ of
eq.(\ref{pic}); in a domain which excludes the sources we have
\begin{equation}\label{dzn}
4\partial_z\partial_{\bar z} \frac{\partial\phi}{\partial z_n} =
e^\phi \frac{\partial\phi}{\partial z_n} 
\end{equation}  
i.e. outside the sources $\frac{\partial \phi}{\partial z_n}$ obeys a
linear elliptic differential 
equation. Similarly we have 
\begin{equation}\label{dgn}
4\partial_z\partial_{\bar z}\frac{\partial\phi}{\partial
g_\infty} = e^\phi 
\frac{\partial\phi}{\partial g_\infty}. 
\end{equation}  
From the previous equations it follows
\begin{equation}\label{green}
\frac{\partial\phi}{\partial z_n}\partial_z\partial_{\bar
z}\frac{\partial\phi}{\partial g_\infty} =  
\frac{\partial\phi}{\partial g_\infty}\partial_z\partial_{\bar
z}\frac{\partial\phi}{\partial z_n}.  
\end{equation} 
If we integrate this expression over a domain $D$ chosen to be
a disk of radius $R$ which includes all the singularities, from which 
disks of radius $\epsilon\rightarrow 0$ around each singularity have
been removed, we can 
apply Green's theorem getting only border contributions.
Letting $R \rightarrow \infty$ and $\epsilon \rightarrow 0$ the only
contributions which survive 
are that coming from the circle at infinity, which gives
\begin{equation}
-\frac{\partial \ln s_\infty}{\partial z_n}
\end{equation}
and the contribution around $z_n$ which gives
\begin{equation}
\frac{\partial \beta_n}{\partial g_\infty}.
\end{equation}
The last equation is due to the fact that one is able to express the
leading term of $\frac{\partial \phi}{\partial z_n}$ and the linear
term in $z-z_n$ of
$\frac{\partial \phi}{\partial g_\infty}$ around $z=z_n$ simply from
the local 
analysis of the solution of the differential equation eq.(\ref{fuchs}). 
Thus one reaches 
\begin{equation}\label{dbetadgdlnsdz}
\frac{\partial \beta_n}{\partial g_\infty} = \frac{\partial \ln
s_\infty}{\partial z_n}. 
\end{equation}
Such a result is sufficient to assure the hamiltonian structure of the
equations eqs.(\ref{equationsofmotion}). In fact taking into account
that $\mu =
2-g_\infty$ we see that the hamiltonian is simply given by $\ln
s_\infty^2$. This result also paves the way to the proof of Polyakov
conjecture. In fact equation (\ref{dbetadgdlnsdz}) can be generalized to
\begin{equation}
\frac{\partial \beta_n}{\partial g_m} = \frac{\partial \ln
s_m}{\partial z_n}. 
\end{equation}
Moreover from the structure of $S_P$ keeping in mind that the
solution of eq.(\ref{pic}) is a stationary point of the action, we
have
\begin{equation}
-\frac{1}{2\pi}\frac{\partial S_P}{\partial g_m} = \ln s_m,
\end{equation}
from which we have
\begin{equation}
-\frac{1}{2\pi} d \frac{\partial S_P}{\partial g_m} = \sum_n
\frac{\partial \beta_n}{\partial g_m} dz_n +{\rm c.c.}
\end{equation}
This is a weak form of Polyakov conjecture which can be rewritten as
\begin{equation}
-\frac{1}{2\pi} d S_P= \sum_n \beta_n dz_n + \sum_n F_n(z_1\dots z_N) dz_n
+{\rm c.c.}
\end{equation}
where the $F_n$ do not depend on the $g_m$.
In order to understand the nature of the functions $F_n$ we consider
the limit of $g_1\rightarrow 0$. In this case $S_P$ becomes
independent of $z_1$ and at the same time $\beta_1 \approx {\rm
const}\times g_1\rightarrow 0$ \cite{hempel}. Thus $F_1$ must be
identically zero. Repeating the reasoning 
for the other singularities we have all the $F_n\equiv 0$.

A more direct proof which exploits the expression of the Polyakov
action in term of a background field has been given in \cite{CMS2}.

\section{Quantization: the two particle case}
 
We shall now quantize the two particle system in the reference
frame which does not rotate at infinity. 
In this case there are no apparent singularities and the hamiltonian is
given by \cite{MS1,CMS} 
\begin{equation}
H = \ln(P z \bar P \bar z) + (\mu -1)\ln (z\bar z)=
\ln(Pz^\mu) + \ln(\bar P\bar z^\mu) = h + \bar h
\end{equation}
with $P = P'_2$ and $z = z'_2$.
$h$ and $\bar h$ are separately constant of motion and if we combine
them with the generalized conservation law \cite{MS1} 
$Pz = (1-\mu)(t-t_0) -ib$ we obtain the solution for the motion 
\begin{equation}
z = {\rm const} [(1-\mu)(t-t_0) - ib]^{\frac{1}{1-\mu}}.
\end{equation}
$H$ can be rewritten as
\begin{equation}
H = \ln((x^2+y^2)^\mu ((P_x)^2 + (P_y)^2)).
\end{equation}    Keeping in mind that with our definitions $P$ is the momentum
multiplied by $16\pi G_N/c^3$, applying the correspondence principle
we have
\begin{equation}
[x,P_x] = [y,P_y] = i l_{P}
\end{equation}
where $l_P = 16 \pi G_n\hbar/c^3$,
all the other commutators equal to zero. $H$ is converted into the
operator
\begin{equation}\label{logbeltrami}
\ln[-(x^2+y^2)^\mu \Delta] +~{\rm constant}.
\end{equation}
The argument of the logarithm is minus the Laplace-Beltrami
$\Delta_{LB}$ operator on the metric $ds^2=(x^2+y^2)^{-\mu} (dx^2 +
dy^2)$. Following
an argument similar to the one presented in
\cite{sorkin} one easily proves that if we start from the
domain of $\Delta_{LB}$ given by the infinite differentiable functions
of compact support $C^\infty_0$ which can also include the origin,
then $\Delta_{LB}$ has a unique
self-adjoint extension in the Hilbert space of functions square
integrable on the metric $ds^2=(x^2+y^2)^{-\mu} (dx^2+dy^2)$ and as a 
result since $-\Delta_{LB}$ is a positive operator, $\ln(-\Delta_{LB})$ is
also self-adjoint. 
 
Deser and Jackiw \cite{deserjackiwcmp} considered the quantum scattering
of a test particle
on a cone both at the relativistic and non relativistic level. Most of
the techniques developed there can be transferred here. The main
difference is the following; instead of the
hamiltonian $(x^2+y^2)^\mu(p_x^2+p_x^2)$ which appears in their non
relativistic treatment, we have now the hamiltonian
$\ln[(x^2+y^2)^\mu(p_x^2+p_y^2)]$. The partial wave eigenvalue
differential equation
\begin{equation}
(r^2)^\mu[-\frac{1}{r} \frac{\partial }{\partial r} r \frac{\partial
}{\partial r}+\frac{n^2}{r^2} ]\phi_n(r) = k^2 \phi_n(r)
\end{equation}
with $\mu=1-\alpha$ is solved by
\begin{equation}
\phi_n(r) = J_\frac{|n|}{\alpha}(\frac{k}{\alpha}r^\alpha)
\end{equation}
to obtain for the Green function \cite{CMS}
$$
G(z,z',t) =
\frac{2}{ \alpha\Gamma(\frac{ict}{l_P})r r'} \left (\frac{r^\alpha
+{r'}^\alpha}{2\alpha}\right )^{2
ict/l_P}
$$
\begin{equation}
\sum_n \frac{e^{in(\phi-\phi')}}{2\pi}
\frac{\Gamma(\frac{|n|}{\alpha}+1 -
\frac{i c t}{l_P})}{\Gamma(\frac{|n|}{\alpha}+1)}
\rho^{\frac{|n|}{\alpha}+1}{_2F_1}(\frac{|n|}{\alpha}+1 - \frac{ict}{l_P};
\frac{|n|}{\alpha}+\frac{1}{2}; 2 \frac{|n|}{\alpha}+1; 4 \rho)\end{equation}
where
\begin{equation}
\rho = \frac{r^\alpha {r'}^\alpha}{r^\alpha+{r'}^\alpha}.
\end{equation}  
The Green function $G$ gives the solution of the two particle quantum
problem.

\end{document}